\documentclass[ap,twocolumn, showkeys,nofootinbib,floatfix]{revtex4-1}
\usepackage{amsmath}
\usepackage{amsfonts}
\usepackage{txfonts}
\usepackage{graphicx}
\usepackage{dcolumn}
\usepackage{bbm}
\usepackage{amssymb}
\usepackage{latexsym}
\usepackage{titlesec}

\begin{document}

\title{ Optimal Photon blockade on the maximal atomic coherence}
\author{Yang Zhang}
\author{Jun Zhang}
\author{Chang-shui Yu}
\email{ quaninformation@sina.com; ycs@dlut.edu.cn}
\affiliation{School of Physics and Optoelectronic Technology, Dalian University of
Technology, Dalian 116024, China }
\begin{abstract}
There is generally no obvious evidence in any direct relation between photon
blockade and atomic coherence. Here instead of only illustrating the photon
statistics, we show an interesting relation between the steady-state photon
blockade and the atomic coherence by designing a weakly driven cavity QED
system with a two-level atom trapped. It is shown for the first time that
the maximal atomic coherence has a perfect correspondence with the optimal
photon blockade. The negative effects of the strong dissipations on photon
statistics, atomic coherence and their correspondence are also addressed.
The numerical simulation is also given to support all of our results.
\end{abstract}
\maketitle
\section{Introduction}

Photon statistics including photon blockade and photon-induced tunneling
have attracted extensive attention in the past years. They result from the
nonlinearity of the cavity field \cite{blocka,hui} and as the
typical nonlinear quantum optical effects, are necessary ingredients for
prospective developments in quantum information processing \cite{optimal}.
Photon blockade indicates the ability to control the nonlinear response of a
system by the injection of single photon \cite{blocka}, while the phenomenon
of photon-induced tunneling is that the system increases the entering
probability of subsequent photons \cite{hui}. These typical features have
been theoretically predicted and experimentally observed in many physical
systems such as opto-mechanical setups, feed-back control system,
super-conducting circuit and so on \cite{Rabl,Nunnenkamp,87,liao,yulong,Hoffman,Liu,Winger,Kimble,77,Verger}.
Cavity quantum electrodynamics (CQED) is an important medium to study the
atom-photon interactions \cite{coherent,kai,Rempe,arka1,arka2,tianlin}. The optical
nonlinearity in such a system arises from the discrete energy level
structure of the atom, so it has an important application in the photon
statistics \cite{Kimble,77,lun,arka,zy}.

Recently, quantum coherence as an essential ingredient of the quantum world
has been widely studied \cite{Adesso1,Adesso,Adesso4,DD,yd,Streltsov,Marvian}. It is at the root of a
number of intriguing phenomena of wide-ranging impact in quantum optics
\cite{Albrecht,Glauber,M}, where decoherence due to the interaction with an
environment is a crucial issue that is of fundamental interest. Quantum
coherence which can also been understood via the theory of physical resource
\cite{plenio1,quant} has attracted increasing interests in many aspects \cite{Yu1,
yao,fan, Lostaglio,Monras} such as hot systems \cite{hot}, many-body systems
\cite{Barontini,Barreiro}, biological system \cite{plenio1,plenio2,Ishizaki,Castro}, low-temperature
 thermodynamics \cite{Lostaglio,berg,Adesso3}, solid-state physics \cite{Lambert}, optimization
of squeezed light \cite{squ} and so on. In particular, one of the intriguing
aspects of quantum coherence is that the atomic coherence has the ability to
enhance the efficiencies of nonlinear optical processes \cite{xm,Harris,Ebert,joo}. Since
the photon blockade requires the strong optical
nonlinearity, is there a clear quantitative relation between the atomic
coherence and photon blockade in a cavity-atom interaction system?
Intuitively, there is no obvious evidence in the relation between the photon
blockade and atomic coherence. So we will turn to another weak question
whether we can find a physical model that can show the relation between
quantum blockade and atomic coherence?

In the present work, we design a weakly driven cavity QED system with a
two-level atom trapped and study the relation between the steady-state
photon blockade and the atomic coherence. Our interest is not to only
illustrate the photon statistics, but to reveal the particularly interesting
correspondence between the photon blockade and the coherence of the atom in
the steady state. As our main result, we find that in the case of steady
state, the atomic coherence has a perfectly consistence with the photon
blockade effect. That is, the maximal atomic coherence just corresponds to
the optimal photon blockade, but the local maximal bunching points which
subject to a two-photon excitation process or a quasi-dark-state process
corresponds to nearly vanishing coherence. In addition, we have also shown
that once the dissipations of the system become relatively strong, the
atomic coherence and the photon anti-bunching effect will be reduced,
meanwhile there appears a deviation in the correspondence between atomic
coherence and photon blockade because of the increasing widths of the energy
levels. The remaining of this paper is organized as follows. In Sec. 2, we
briefly describe our model. In Sec. 3, we discuss the mechanism of photon
-induced tunneling and photon blockade by the analytical method. In Sec. 4,
we study the atomic coherence and present the correspondence relation
between coherence and different statistics, whilst a numerical simulation
based on the master equation is also given to support our analytic
treatment. Finally, we draw our conclusion and give some discussions.

\section{ The physical model}
\begin{figure}[tbp]
\centering
\hspace*{-2.5cm} \includegraphics[width=1.5\columnwidth,height=3in]{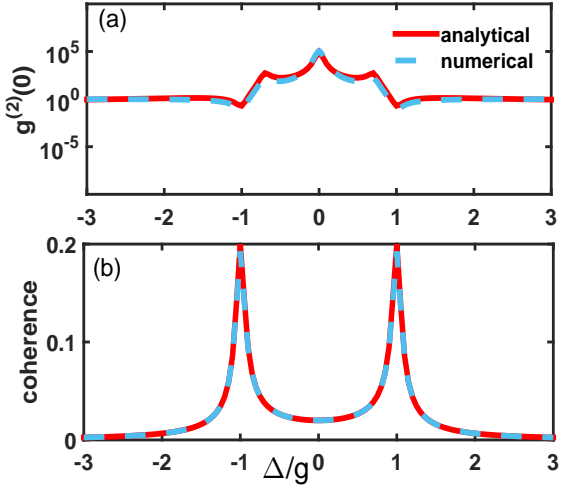}
\caption{(color online). The equal-time second-order function $g^{(2)}(0)$
and the coherence of atom vs the detuning $\Delta $, respectively. The red
curves are analytical results and the green curves are numerical results of
the quantum master equation. We take $\protect\gamma /g=0.05,$ $\protect%
\kappa /g=0.05$, $\protect\eta /g=0.01$. }
\end{figure}
The system we studied here consists of a single atom (with ground state $%
\left\vert g\right\rangle $ and excited state $\left\vert e\right\rangle $)
coupled to a cavity mode and the cavity is weakly driven by a laser with the
frequency denoted by $\omega _{L}$ and the Rabi frequency denoted by $\eta $%
. The coupled system is well governed by the Hamiltonian $H$ (we set $\hbar
=1$ hereafter)
\begin{eqnarray}
H &=&\omega _{a}a^{\dag }a+\omega _{e}\sigma ^{+}\sigma ^{-}+g(\sigma
^{+}a+a^{\dag }\sigma ^{-})  \label{(aa)} \\
&&+\eta \left( a^{\dag }e^{-i\omega _{L}t}+ae^{i\omega _{L}t}\right),
\nonumber
\end{eqnarray}%
where $a$ and $\sigma ^{-}=\left\vert g\right\rangle \left\langle
e\right\vert $ are the annihilation and lowering operators for the cavity
mode and the atom, respectively and $\omega _{e}$ is the frequency of atomic
transition from ground state $\left\vert g\right\rangle $ to excited state $%
\left\vert e\right\rangle $  and $\omega _{a}$ is
the cavity resonance frequency. In addition, we set the coupling coefficient
between the atom and the cavity mode to be $g$. In the frame rotated at the
laser frequency $\omega _{L}$, the Hamiltonian (\ref{(aa)}) can be rewritten
as
\begin{eqnarray}
H &=&\Delta _{a}a^{\dag }a+\delta \sigma ^{+}\sigma ^{-}+g(\sigma
^{+}a+a^{\dag }\sigma ^{-})  \nonumber \\
&&+\eta \left( a^{\dag }+a\right),  \label{bbb}
\end{eqnarray}%
with the laser detuning from the cavity mode $\Delta _{a}=\omega _{a}-\omega
_{L}$ and $\allowbreak \delta =\omega _{e}-\omega _{L}$ corresponding to the
laser detuning from the atom.

For simplicity, we assume that the cavity is resonant with the atom, i.e., $%
\omega _{a}=\omega _{c}$ and $\Delta _{a} =\delta =\Delta$. Since the system
is driven weakly, only few photons can be excited. Thus we can only focus on
the few-photon subspace. In this regime, energy eigenstates are in two-level
manifolds. So the eigenenergies and eigenstates of the Hamiltonian without
driving can be given by
\begin{eqnarray}
E_{n\pm } &=&n\Delta \pm g\sqrt{n},  \nonumber \\
\left\vert n,-\right\rangle &=&\frac{1}{\sqrt{2}}(\left\vert
n,g\right\rangle -\left\vert n-1,e\right\rangle ), \nonumber \\
\left\vert n,+\right\rangle &=&\frac{1}{\sqrt{2}}(\left\vert
n,g\right\rangle +\left\vert n-1,e\right\rangle ),
\end{eqnarray}%
where $n$ is the number of energy quanta in the CQED system (in the weak
driving regime, we can safely cut off the photon number to 2 ) which
distinguishes the different eigenstates. One can easily find the
eigenenergies in the manifolds depend on the coupling $g$ (the equivalent
coupling rate is $g\sqrt{n}$). The nonlinearity in the coupling between the
atom and cavity gives rise to energy level structure which can exhibit
different photon statistics behaviors due to the splitting of the
eigenenergy \cite{jc}.

When the environmental effect is taken into account in the current system,
there are two mechanisms for energy dissipation: cavity decay and
spontaneous emission of the atom. We assume zero temperature reservoirs, the
corresponding master equation can be given in the following Lindblad form
\begin{eqnarray}
\dot{\rho} &=&-i[H,\rho ]+\kappa (2a\rho a^{\dagger }-a^{\dagger }a\rho
-\rho a^{\dagger }a)  \nonumber \\
&+&\gamma (2\sigma ^{-}\rho \sigma ^{+}-\sigma ^{+}\sigma ^{-}\rho -\rho
\sigma ^{+}\sigma ^{-}),  \label{mast}
\end{eqnarray}%
where $H$ is the system's original Hamiltonian as given by Eq. (\ref{bbb}),
$\rho$ is the density operator for the atom-cavity system, $\kappa $ is the
field decay rate for the cavity mode, $\gamma $ is the atomic spontaneous
emission rate. In general, for a full quantum mechanical treatment of the
system, we can compute the numerical solutions to the master equation Eq. (%
\ref{mast}) using truncated number state bases for the cavity mode \cite{77}%
. Here, we restrict ourselves in the subspace spanned by the basis $%
\{\left\vert n,g\right\rangle ,$ $\left\vert n-1,e\right\rangle \},$ hence
the formal solution of $\rho $ can be gotten. Once $\rho $ is given, any
physical quantity of the system can be obtained.

\section{ The photon statistics}

As mentioned at the beginning, the signatures of the photon behaviors can be
detected through photon statistics measurement \cite{arka},\ which can be
characterized by the normalized the equal-time correlation function, which
is defined for stationary state \cite{scully}

\begin{equation}
g^{(2)}(0)=\frac{\left\langle a^{\dagger }a^{\dagger }aa\right\rangle }{%
\langle a^{\dagger }a\rangle ^{2}}=\frac{Tr[\rho _{s}a^{\dagger 2}a^{2}]}{%
[Tr(\rho _{s}a^{\dagger }a)]^{2}},  \label{nn}
\end{equation}%
where $a$ is the annihilation operator for the cavity mode, the $\rho _{s}$
is the steady-state density matrix of the composite system which can be
obtained by employing a numerical way to solving the master equation in Eq. (%
\ref{nn}) \cite{tan1}. The photon blockade which means the system 'blocks'
the absorption of a second photon with the same energy and large
probability. The limit $g^{(2)}(0)\rightarrow 0$ means the perfect photon
blockade in which two photons never occupy the cavity at the same time. On
the contrary, when $g^{(2)}(0)>1$, it means photons inside the cavity
enhance the resonantly entering probability of subsequent photons \cite{li1,li2,lang}.

In order to gain more insight into the physics, we first take an analytic
(but approximate) method to calculate the second-order correlation function
with the help of the wave function amplitude approach by employing the Schr%
\"{o}dinger equation \cite{open}
\begin{equation}
i\frac{d\left\vert \Psi \right\rangle }{dt}=H_{eff}\left\vert \Psi
\right\rangle.
\end{equation}%
Considering the effects of the two channels (the leakage of the cavity $%
\kappa $, the spontaneous emission $\gamma $ of the atom) \cite{}, we
phenomenologically add the relevant damping contributions to Eq. (\ref{bbb}%
). Thus the effective Hamiltonian can be rewritten as
\begin{equation}
H_{eff}=H-\frac{i}{2}(\kappa a^{\dagger }a+\gamma \sigma ^{+}\sigma ^{-}).
\end{equation}%
Analogously to the above statements, we can omit the probability for three
or more photons due to the weak driving. This makes it easy to evaluate
following analytical expression. In this case, we can suppose that the state
of the system can be expressed as \cite{fliuds,steady1,Carmichael}
\begin{eqnarray}
\left\vert \Psi \right\rangle &=&C_{0g}\left\vert 0,g\right\rangle
+C_{1g}\left\vert 1,g\right\rangle +C_{0e}\left\vert 0,e\right\rangle
\nonumber \\
&&+C_{2g}\left\vert 2,g\right\rangle +C_{1e}\left\vert 1,e\right\rangle.
\label{phi}
\end{eqnarray}%
\begin{figure}[tbp]
\centering
\hspace*{-1cm} \includegraphics[width=1.3%
\columnwidth,height=2in]{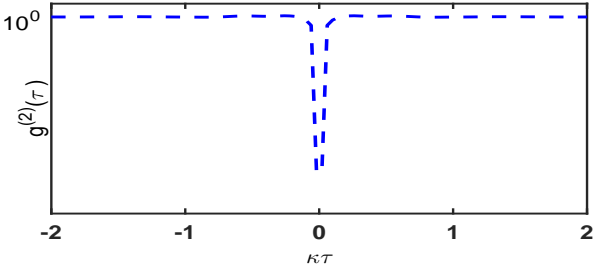}
\caption{(color online). The evolution of finite time delay correlation
function $\ g^{(2)}(\protect\tau )$ of the cavity mode by numerical
simulation the master equation. $\ g^{(2)}(\protect\tau )$ exhibits photon
antibunching $g^{(2)}(0)\prec g^{(2)}(\protect\tau )$, and sub-Poissonian
photon statistics $g^{(2)}(0)\prec 1$. Here, $\protect\gamma /\protect\kappa
$=1, $\Delta /\protect\kappa =\protect\delta /\protect\kappa=-20,$ $g/%
\protect\kappa=20$ . }
\end{figure}
Based on the definition of $g^{(2)}(0)$, i.e.,
\begin{equation}
g^{(2)}(0)=\frac{\sum_{n}n(n-1)p_{n}}{(\sum_{n}np_{n})^{2}}=\frac{2p_{2}}{%
(p_{1}+2p_{2})^{2}},
\end{equation}%
where $p_{n}=$ $\left\vert C_{n}\right\vert ^{2}$ represents the probability
with $n$ photons and $p_{1}=\left\vert \bar{C}_{1g}\right\vert
^{2},p_{2}=\left\vert \bar{C}_{2g}\right\vert ^{2}$, one find that $%
g^{(2)}(0)$ can be obtained so long as one can solve $C_{n}$ given in Eq. (%
\ref{phi}). To do so, we substitute $\left\vert \Psi \right\rangle $ into
Eq. (6) and arrive at the following dynamical equations
\begin{eqnarray}
\dot{C}_{1g} &=&-(\kappa /2+i\Delta _{a})C_{1g}-i\eta C_{0g}-igC_{0e}-\sqrt{2%
}i\eta C_{2g},  \nonumber \\
\dot{C}_{0e} &=&-(\gamma /2+i\delta )C_{0e}-igC_{1g}-i\eta C_{1e},  \nonumber
\\
\dot{C}_{2g} &=&-2(\kappa /2+i\Delta _{a})C_{2g}-\sqrt{2}igC_{1e},
\label{steady} \\
\dot{C}_{1e} &=&-(\kappa /2+\gamma /2+i\delta +i\Delta _{a})C_{1e}-\sqrt{2}%
igC_{2g}-i\eta C_{1e}.  \nonumber
\end{eqnarray}
Let the initial state of the system be $\left\vert 0,g\right\rangle $.
Considering the limit of the weakly driving field again, we can get $\bar{C}%
_{0g}$ $\rightarrow $ $1$, and Eq. (\ref{steady}) are closed. Thus, Eq. (\ref%
{steady}) can be easily solved. In the following, we will only consider the
question in the steady-state case. In addition, the steady-state solution of
Eq. (\ref{steady}) can be analytically obtained, but the precise form are
quite cumbersome, so we further neglect the high-order terms of $\eta $
(weak driving) in Eq. (\ref{steady}). Under these conditions, we can get the
steady-state solution of Eq. (\ref{steady}) as follows.
\begin{eqnarray}
&&\bar{C}_{1g}=-\frac{i\eta \alpha }{g^{2}+\alpha \beta }, \\
&&\bar{C}_{0e}=-\frac{g\eta }{g^{2}+\alpha \beta }, \\
&&\bar{C}_{2g}=-\frac{\eta ^{2}[g^{2}-\alpha ^{2}-\alpha \beta ]}{\sqrt{2}%
(g^{2}+\alpha \beta )(g^{2}+\beta ^{2}+\alpha \beta )}, \\
&&\bar{C}_{1e}=\frac{ig^{2}\eta ^{2}(\alpha +\beta )}{(g^{2}+\alpha \beta
)(g^{2}+\beta ^{2}+\alpha \beta )},
\end{eqnarray}%
with $\alpha =(\gamma /2+i\delta ),$ $\beta =(\kappa /2+i\Delta _{a}).$ In
the weak-driving case, we can easily get $p_{1}\gg p_{2}$, then the
equal-time second-order correlation function can be simplified as $%
g^{(2)}(0)\approx \frac{2p_{2}}{(p_{1})^{2}}$. It can be further written by
\begin{equation}
g^{(2)}(0)\approx \frac{xy}{z},  \label{aa}
\end{equation}%
where $x:=(g^{2}+\alpha \beta )(g^{2}+\alpha ^{\ast }\beta ^{\ast }),$ $%
y:=(g^{2}-\alpha (\alpha +\beta ))(g^{2}-\alpha ^{\ast }(\alpha ^{\ast
}+\beta ^{\ast })),$ $z:=\alpha ^{2}\alpha ^{\ast 2}(g^{2}+\beta (\alpha
+\beta ))(g^{2}+\beta ^{\ast }(\alpha ^{\ast }+\beta ^{\ast })).$

In Fig. 1 (a), we plot $g^{(2)}(0)$ with the detuning $\Delta $ to
illustrate the behaviors of photon antibunching and bunching in the case of
weak dissipations and driving (here we let $\Delta _{a}=\delta =$ $\Delta $%
). In this figure, $\Delta =\{\pm g,\pm \sqrt{2}g/2\}$ corresponds to the
photon sub-Poissonian and super-Poissonian, respectively. In the blockade
regime, the successful blocking of the second photon depends on how well the
first photon is coupled to the CQED system \cite{arka1}. The quantum
signatures can also be manifested by a finite time delay second-order
function $g^{(2)}(\tau ),$ Based on the second-order correlation function,
one can know whether the photon anti-bunching happened or not. The photon
antibunching can be demonstrated by a rise of $g^{(2)}(\tau )$ with $\tau $
increasing from 0 to larger values while $g^{(2)}(0)\prec g^{(2)}(\tau )$
\cite{pp,lang}. Since reaching $g^{(2)}(\tau )\succ $ $g^{(2)}(0),$ it
violates Cauchy-Shwarz inequality and is a nonclassical effect. As shown in
Fig. 2, photon antibunching can be observed, $g^{(2)}(0)$ is at a minimum
and $g^{(2)}(\tau )$ rise for increasing $\tau .$

\section{Atomic coherence and photon statistics}
\begin{figure}[tbp]
\centering
\hspace*{-1cm}\includegraphics[width=1.3%
\columnwidth,height=2.5in]{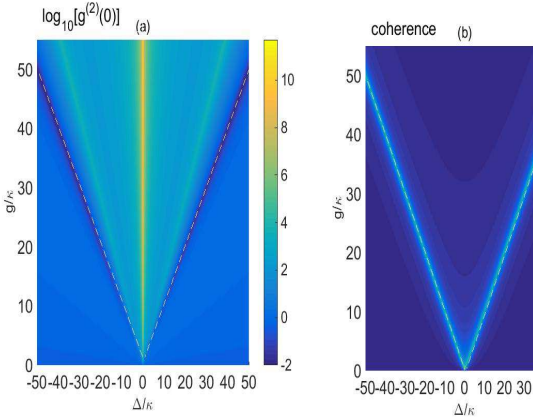}
\caption{(color online). (a) Logarithmic plot the equal-time second-order
function $g^{(2)}(0)$ in (a) as a function of the cavity-atom coupling rate $%
g$ and detuning $\Delta $. (b) The plot of the atomic coherence varying with
$g $ and $\Delta $. The optimal photon anti-bunching and the maximal
coherence are illustrated by the white-dashed line respectively in (a) and
(b) which corresponds to $\Delta ^{2}=g^{2}$. Here, $\protect\gamma /\protect%
\kappa $=0.5, $\protect\eta /\protect\kappa =0.1$.}
\end{figure}

In this section, we will give a detailed investigation of the atom's
coherence. It has been shown that the off-diagonal elements of $\rho $
characterize interference. They are usually called as coherence with respect
to the basis in which $\rho $ is written \cite{wall,Yu,ot,ann}. The
coherence can be measured by \cite{Yu}
\begin{equation}
C(\rho )=\left\Vert \rho -\sigma ^{\ast }\right\Vert _{1}=\sum\limits_{i\neq
j}\left\vert \rho _{ij}\right\vert,  \label{coherence}
\end{equation}%
where $\left\Vert .\right\Vert _{1}$ is $l_1$ norm and $\sigma ^{\ast }$
denotes the diagonal matrix with $\sigma _{ij}^{\ast }=\rho _{ij}.$ It is
shown that the coherence measure has a direct geometric meaning and this
measure removes the all-diagonal elements and collects the contribution of
off-diagonal elements of $\rho $.

Since we have obtained the steady solutions of Eq. (\ref{steady}), we can
calculate the state $\left\vert \Psi \right\rangle $ given in Eq. (\ref{phi}%
) and the reduced density matrix $\rho _{A}$ for the atom. Thus based on the
above method, we can naturally calculate the corresponding coherence. So
from the state $\left\vert \Psi \right\rangle $, one can find that the
reduced density matrix of the atom is
\begin{equation}
\rho _{A}=Tr_{R}\left\vert \Psi \right\rangle \left\langle \Psi \right\vert
=\left\vert \psi _{0}\right\rangle \left\langle \psi _{0}\right\vert
+\left\vert \psi _{1}\right\rangle \left\langle \psi _{1}\right\vert
+\left\vert \bar{C}_{2g}\right\vert ^{2}\left\vert g\right\rangle
\left\langle g\right\vert ,
\end{equation}%
with
\begin{equation}
\left\vert \psi _{0}\right\rangle =\bar{C}_{0g}\left\vert g\right\rangle +%
\bar{C}_{0e}\left\vert e\right\rangle,
\end{equation}%
\begin{equation}
\left\vert \psi _{1}\right\rangle =\bar{C}_{1g}\left\vert g\right\rangle +%
\bar{C}_{1e}\left\vert e\right\rangle,
\end{equation}%
and the subscript $R$ means trace over cavity field. Thus one can arrive at
\begin{eqnarray}
&&\rho _{A}=(\left\vert \bar{C}_{0g}\right\vert ^{2}+\left\vert \bar{C}%
_{1g}\right\vert ^{2}+\left\vert \bar{C}_{2g}\right\vert ^{2})\left\vert
g\right\rangle \left\langle g\right\vert +  \label{cc} \\
&&(\bar{C}_{0g}\bar{C}_{0e}^{\ast }+\bar{C}_{1g}\bar{C}_{1e}^{\ast
})\left\vert g\right\rangle \left\langle e\right\vert +(\bar{C}_{0e}\bar{C}%
_{0g}^{\ast }+\bar{C}_{1e}\bar{C}_{1g}^{\ast })\left\vert e\right\rangle
\left\langle g\right\vert   \nonumber \\
&&+(\bar{C}_{0e}\bar{C}_{0e}^{\ast }+\bar{C}_{1e}\bar{C}_{1e}^{\ast
})\left\vert e\right\rangle \left\langle e\right\vert .  \nonumber
\end{eqnarray}%
\newline
Since $\bar{C}_{0g}\rightarrow 1$, $\left\vert \bar{C}_{0g}\right\vert
^{2}\sim \eta ^{0},$ $(\bar{C}_{0g}\bar{C}_{0e}^{\ast },\bar{C}_{0e}\bar{C}%
_{0g}^{\ast })\sim \eta $, $(\left\vert \bar{C}_{1g}\right\vert ^{2},\bar{C}%
_{0e}\bar{C}_{0e}^{\ast })\sim \eta ^{2}$, $(\bar{C}_{1g}\bar{C}_{1e}^{\ast
},\bar{C}_{1e}\bar{C}_{1g}^{\ast })\sim \eta ^{3}$ and $(\left\vert \bar{C}%
_{2g}\right\vert ^{2},\bar{C}_{1e}\bar{C}_{1e}^{\ast })\sim \eta ^{4},$ the
elements of the $\rho _{A}$ can be expanded based on the small $\eta $. To
proceed, we can find that the anti-diagonal element is well determined by
the function of $\eta $: $f(\eta )=a+b\eta +c\eta ^{2}+...$.. Interestingly,
to a good approximation, and basing on the steady amplitudes derived from
Eq. (6), the coherence is analytically given by
\begin{equation}
C(\rho _{A})=\frac{2g\eta }{\sqrt{x}}.  \label{bb}
\end{equation}%

Next, we will discuss the relations between the coherence and photon
blockade in the CQED system. We have plotted Eq. (\ref{bb}) in Fig. 1 (b).
We also plot the coherence via the numerical way by solving the master
equation in this figure. In the analytic case, the total number of photon is
truncated at $n\preceq 2$ to the weak driving. However, we use the numerical
calculation to check the analytic results by the master equation where the
dimension of photon space is approximately truncated to 4. We also use
higher dimensional photon space, no obvious difference appears. In this
sense, we think the current dimension of the space is acceptable. One can
find that the analytical results matches the numerical simulations very
well, which shows the validity of our approximate and analytic results.
\begin{figure}[tbp]
\centering
\hspace*{-2cm}\includegraphics[width=1.45%
\columnwidth,height=3.in]{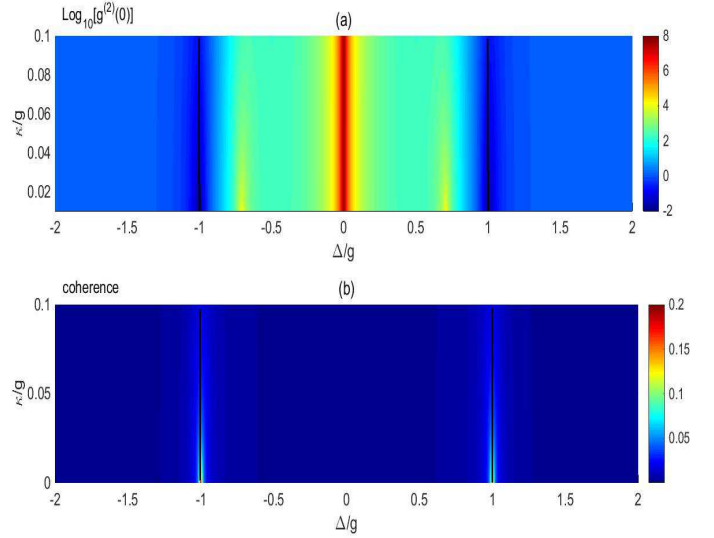}
\caption{(color online). We plot logarithm of the equal-time second-order
correlation function $\log _{10}\left[ g^{(2)}(0)\right] $ and the atomic
coherence as a function of the detuning $\Delta $ and cavity decay rate $%
\protect\kappa $. (a) shows the second-order correlation function and (b)
corresponds the atomic coherence. The optimal photon anti-bunching and the
maximal atomic coherence are also illustrated by the black-solid line in (a)
and (b) which corresponds to $\Delta^2=g^2$. Here, we set $\protect\gamma%
/g=0.01 $, $\protect\eta/g=0.001$.}
\end{figure}
From Fig. 1 (b), it is obvious that the coherence has a pair of maximal
values. Compared with Fig. 1 (a), one can easily find that the two points
with the maximal coherence perfectly correspond to the optimal photon
blockade, which means the maximal coherence of the atomic state can capture
the photon blockade. In addition, one can also see that the maximal photon
bunching point corresponds to the almost vanishing coherence instead of the
maximal coherence. To better understand this correspondence relation, let us
use the analytic expression given in Eq. (\ref{bb}) to analyze the results,
from which one can see that the extremum occur at $\Delta =\pm g$ for small $%
\{\kappa ,\gamma \}$. This is consistent with the analysis on the optimal
photon anti-bunching condition. In order to give an intuitive understanding
of this correspondence, one should note that once $\Delta =\pm g$, the
driving field is tuned resonantly with the transition between $\left\vert
0g\right\rangle $ and $\left\vert 1,\pm \right\rangle $ which leads to the
optimal photon blockade, meanwhile, the system will locate in the state $%
\bar{C}_{1g}\left\vert 1,g\right\rangle +\bar{C}_{0e}\left\vert
0,e\right\rangle $. Thus, the state of $\left\vert 0,e\right\rangle $ is
actually occupied with a relatively large probability which is proportional
to the first order of the driving field $\eta $. It is obvious from Eq. (\ref%
{bb}) that $\left\vert \Psi \right\rangle $ owns the relatively large amount
of coherence (see in Fig. 1 (b)). When the driving field is resonant to the
two-photon process (the transition $\left\vert 0,g\right\rangle $ to $%
\left\vert 2,\pm \right\rangle )$, thus $\left\vert 2,\pm \right\rangle $
occupies the relatively dominant proportion in $\left\vert \Psi
\right\rangle $. However, the amplitude is proportional to the $\eta ^{2}$.
To the good approximation, we can omit it safely in Eq. (\ref{cc}). Thus,
the coherence at these points do not get the extremum. In addition, it is
interesting that the maximal photon bunching point at $\Delta =0$
(photon-induced tunnelling point) does not correspond to an extremum of
coherence. At $\Delta =0,$ a quasi-dark state process occurs, the system
being driven into a quasi-dark state $\left\vert d\right\rangle \thicksim
g\left\vert 0,g\right\rangle -\eta \left\vert 0,e\right\rangle ,$ which
provides a channel to be converted to the state $\left\vert 2,g\right\rangle
$ as well as $\left\vert 1,e\right\rangle $. Their proportions in $%
\left\vert \Psi \right\rangle $ get relatively larger. The net effect on
coherence is that $\left\vert 0,e\right\rangle $ reaches a suppression in
the driven mode, so the coherence is negligibly small.

In addition, in order to find the general features of the correspondence
relation, we plot the coherence and the logarithmic equal-time second-order
correlation function with different $\Delta $ and $g$ in Fig. 3. From the
Fig. 3, we can observe that optimal anti-bunching corresponds to the maximal
coherence with small enough dissipations ($g,\Delta \gg \{\kappa ,\gamma \}$%
) which are plotted by the white dashed lines in the Fig. 3 (a) and Fig. 3
(b). The condition for this correspondence relation can be demonstrated
analytically from Eq. (\ref{aa}) and Eq. (\ref{bb}) as $\Delta ^{2}=g^{2}.$
In addition, one can find that the dissipations of the cavity and atom have
certain influence on both photon statistics and atomic coherence. We plot
the effect of the cavity decay $\kappa $ on the correspondence relation in
Fig. 4. We note that both the optimal photon statistics and the coherence
extremals are reduced with $\kappa $ increasing. Meanwhile, the
correspondence relation between coherence and $g^{2}(0)$ gets worse.
Mathematically, this can be well understood from Eq. (\ref{aa}) and Eq. (\ref%
{bb}) from which one can see that all the relevant analysis are satisfied
within the error region to the same order as $\kappa ^{2}$. Physically, the
inaccuracy obviously results from the increasing widths of the energy
levels. So generally we always limit our study in the region with small
enough dissipations for a good correspondence relation.

\section{Discussions and conclusion}

To conclude, we have studied the photon statistics and the atomic coherence
in a weakly driving CQED system and analyzed the physical mechanisms of
photon statistics and coherence in details. We systematically study the
system parameters' dependence on the photon statistics and coherence. Our
results gave a clear quantitative analysis and connections between atomic
coherence and quantum statistics. By numerically solving the master equation
in the steady state and calculating the equal time second-order correlation
function and the atomic coherence, we obtained a perfect relation between
the atomic coherence and the photon blockade, i.e., the maximal coherence
always correspond to the optimal anti-bunching points. By the analytical
way, we derive the analytical condition for the correspondence relation,
which agrees well with the numerical simulation. In addition, the maximal
photon bunching point corresponds to the almost vanishing coherence due to a
quasi-dark-state process.

Finally, we would like to emphasize that this perfect match relation has
been only found in the current special model after all. Whether this
relation can exist in other model with much strong quantum blockade deserves
us forthcoming research. If such a model could be found, one maybe could
enhance the photon blockade by controlling the atomic coherence. In summary,
we provided new insight into the rather unexplored area and could be
potentially pivotal for future quantum applications.

\section{Acknowledgements}
Y Zhang  thanks J. S. Jin for valuable discussion. This work was supported by
the National Natural Science Foundation of China, under Grant No.11375036
and 11175033, the Xinghai Scholar Cultivation Plan and the Fundamental
Research Funds for the Central Universities under Grants No. DUT15LK35 and
No. DUT15TD47.

\end{document}